# Superconductor made by electrolyzed and oxidized water

Chia-Jyi Liu[1], Tsung-Hsien Wu[1], Lin-Li Hsu[1], Jung-Sheng Wang[1], Shu-Yo Chen[1], Wei Jen Chang[2], & Jiunn-Yuan Lin[3]

[1]*Department of Physics, National Changhua University of Education, Changhua, 50007, Taiwan*

[2]*Department of Electrophysics, National Chiao tung University, Hsinchu 30050, Taiwan*

[3]*Institute of Physics, National Chiao tung University, Hsinchu 30050, Taiwan*

By deintercalation of $Na^+$ followed by inserting bilayers of water molecules into the host lattice, the layered cobalt oxide of gamma-$Na_{0.7}CoO_2$ undergoes a topotactic transformation to a layered cobalt oxyhydrate of $Na_{0.35}(H_2O)_{1.3}CoO_{2-delta}$ with the *c*-axis expanded from c = 10.9 anstrom to c = 19.6 anstrom. In this paper, we demonstrate that the superconducting phase of c = 19.6 anstrom can be directly obtained by simply immersing gamma-$Na_{0.7}CoO_2$ powders in electrolyzed/oxidized (EO) water, which is readily available from a commercial electrolyzed water generator. We found that high oxidation-reduction potential of EO water drives the oxidation of the cobalt ions accompanying by the formation of the superconductive c = 19.6 anstrom phase. Our results demonstrate how EO water can be used to oxidize the cobalt ions and hence form superconducting cobalt oxyhydrates in a clean and simple way and may provide an economic and environment-friendly route to oxidize the transition metal of complex metal oxides.

Superconducting cobalt oxyhydrate $Na_{0.35}(H_2O)_{1.3}CoO_{2-\delta}$ with $T_c \approx 4.5$ K can be synthesized either via a simple process by immersing it in an oxidizing solution such as $Br_2$ in $CH_3CN$[1], $Na_2S_2O_{8(aq)}$[2], $KMnO_{4(aq)}$[3], and $NaMnO_{4(aq)}$[4], or anodically polarizing it [5] with a constant voltage of 0.6-1.2 V at room temperature. Like high-$T_c$ copper oxide superconductor, the superconducting transition temperature $T_c$ and the oxidation number of cobalt ion are associated with the cationic content in the charge reservoir and the oxygen content in the cobalt oxyhydrates.[6,7] During electrolysis of water, it creates on the anode area an acidic water, which is severely lacking electrons and is therefore called electrolyzed/oxidized (EO) water having a high oxidation-reduction potential (ORP). We have found that simply immersing 100 mg of $\gamma$-$Na_{0.7}CoO_2$ powders in the 250 ml EO water can result in the same superconductive $c \approx 19.6$ Å phase of $Na_{0.35}(H_2O)_{1.3}CoO_{2-\delta}$ with $T_c \approx 4$ K. This is achieved by the deintercalation of $Na^+$, oxidation of Co ions and subsequent hydration process of intercalating the water molecules between the $CoO_2$ layers and $Na^+$ layers. Figure 1 shows the powder x-ray diffraction pattern for the EO water-treated sample. All the reflections of the EO water-treated product are the same as those of the $Br_2$- or $NaMnO_4$- treated samples, which indicates the same crystal structure having space group $P6_3/mmc$ (space group # 194) with $c \approx 19.6$ Å evidenced by the fingerprint of (002) reflection occurring at $2\theta \approx 11.3°$ (d spacing $\approx 9.81$ Å) for Fe $K_\alpha$ radiation.

We now know that the superconductive $c \approx 19.6$ Å phase is unstable in the ambient environment and readily transforms into $c \approx 13.8$ Å phase in an environment without sufficient humidity.[8] Figure 2 shows that the (002) reflection at $2\theta \approx 11.3°$ of the EO-treated sample still remains after being exposed to ambient air for 21 days, which is in sharp contrast with the instability of the Br-treated $Na_{0.35}(H_2O)_{1.3}CoO_{2-\delta}$. One can see the (004) reflection turns into a broader and asymmetric peak, which is likely associated with the initial stage of disorder in the interlayer distance of $CoO_2$ layers caused by the random stacking of water bilayers and monolayers between the $CoO_2$ layers, a reminiscent case of $Na_{1/3}(H_2O)_{1.5}TaS_2$.[9]

From all the previous reports, it is also well known that the fully hydrated c ≈ 19.6 Å phase is a superconductor with $T_c$ ≈ 4 K. In order to confirm the superconductivity of the EO-treated sample, Fig. 3 shows the dc mass magnetization data of the EO water-treated sample measured in a field of 10 Oe down to 2 K. Both zero field cooled and field cooled data clearly show a diamagnetic signal of magnetization and a rapid decrease starting with 4.2 K, indicating a superconducting transition with an onset temperature of 4.2 K. The mass magnetization at 2 K is $-3.47 \times 10^{-3}$ emu/g and $-7.49 \times 10^{-3}$ emu/g in the field-cooled mode and zero-field- cooled mode, respectively.

Furthermore, we have also found that simply immersing γ-$Na_{0.7}CoO_2$ in the tap water without using any other oxidizing agent can result in a deep de-intercalation of $Na^+$ and large oxygen deficiency but would not oxidize the Co ions. The composition analysis using inductively coupled plasma atomic-emission spectrometry (ICP-AES), water content determination using thermogravimetric analyzer (TGA), and oxygen content determination using iodometric titration show that the tap water-treated product has the chemical formula of $Na_{0.29}(H_2O)_{0.41}CoO_{1.78}$. The spontaneous deintercalation of $Na^+$ from the host lattice might be due to the stronger Na-Co Coulomb repulsion for $Na^+$ in the *2b* site as compared that in the *2d* site, which makes γ-$Na_{0.7}CoO_2$ have the escape tendency of $Na^+$ when being immersed in a solution. After deintercalating some of the $Na^+$ from the host lattice, the coulomb attraction between the positive charges of $Na^+$ and the negative charges of $O^{2-}$ becomes smaller and consequently results in a slight expansion along the *c*-axis. This fact can be seen in Fig. 2. As compared to 2θ = 20.44° (*d* = 5.456 Å) in the XRD pattern for γ-$Na_{0.7}CoO_2$, the (002) reflection of the tape water-treated sample is found to shift to a lower angle at 2θ = 20.02° (*d* = 5.569 Å), which corresponds to c ≈ 11.2 Å. One should note that treating γ-$Na_{0.7}CoO_2$ using the EO water having insufficient ORP leads to a similar XRD pattern to that using the tap water.

The energy shifts of *d* orbital in transition metals arising from metal-anion ligand interactions could be described by the angular overlap model (AOM),[10] which is a first approximation to the full molecular orbital model and could facilitate the understanding of the principles of transition metal oxides. For octahedral geometry, *d* orbitals are split into threefold $t_{2g}$ and twofold $e_g$ states by the crystal field. For the $Na_xCoO_{2-\delta}$ system, the threefold degenerate $Co^{3+/4+}$:$t_{2g}$ band is further split by the trigonal crystal field into an $a_{1g}$ band and a twofold degenerate $e_g'$ band. Due to the covalent mixing of the Co *3d* and O *2p* states, electrons can be removed either from the Co *3d* or O *2p* bands upon oxidation of the cobaltates, which depends on which is the highest occupied molecular orbital (HOMO). Based on the wet-chemical redox analyses, $Na_xCoO_{2-\delta}$ and $Li_{1-x}CoO_{2-\delta}$ share some similarity at deep de-intercalation of alkaline metal.[7,11] The oxidation number of Co for both systems reaches a nearly constant value at deep de-intercalation of alkaline metal and could not increase further. Instead, large oxygen deficiency is always found in these materials to compensate for the lower-than-expected oxidation number of Co, which is presumably a process of removing electrons from the O *2p* bands instead of Co *3d* bands to make up the loss of $Na^+$ from the host lattice. Since the EO-treated sample has the same superconductive c ≈ 19.6 Å phase, we use $Na_{0.36}(H_2O)_{1.3}CoO_{1.91}$ in ref. 12 to compare with $Na_{0.29}(H_2O)_{0.41}CoO_{1.78}$ with c ≈ 11.2 Å regarding the oxidation number of Co and oxygen deficiency. The oxidation number of Co is +3.48 and +3.27 for the former and the latter, respectively. One can readily see the significant difference in the oxidation number of Co at deep deintercalation of $Na^+$, which could be associated with the ORP of the solution used to treat the parent material γ-$Na_{0.7}CoO_2$. The EO water has the ORP of ~880 mV, whereas the tap water has the ORP of ~600 mV. The Hanna 7022 ORP solution with ORP = 470 mV at 25°C is used to calibrate the platinum electrode for measuring the ORP of EO water before use. It seems that to remove electrons from the $Co^{3+/4+}$:$a_{1g}$ band and oxidize the Co ions requires an oxidizing solution with sufficient ORP. Insufficient ORP solution would not oxidize the Co ions and only results in oxygen deficiency to maintain the charge neutrality at deep deintercalation of $Na^+$. It is conceivable that hydrogen bonding between oxygen of the $CoO_2$ layers and hydrogen of the water molecules play a significant role in stabilizing the c ≈ 19.6 Å phase.[13] Therefore, the oxygen deficiency in the tape water-treated sample could be responsible for being unable to form the c ≈ 19.6 Å phase due to insufficient amount of hydrogen bonding.

In spite that some consider the electrolyzed water as aqua scam, a few benefits are claimed for using electrolyzed water in dentistry, antiseptic, and horticulture applications. This work reveals several experimental findings that are critical for understanding the phase formation of superconductive sodium cobalt oxyhydrates. Sufficient ORP of the EO water seems to be essential to transform the γ-$Na_{0.7}CoO_2$ to the c ≈ 19.6 Å phase. The key factor of forming the c ≈ 19.6 Å phase relies on whether the oxidizing solution has the ability to abstract electrons from the cobalt ions. Deintercalation of $Na^+$ for γ-$Na_{0.7}CoO_2$ is spontaneous and would not necessarily bring about oxidation of the cobalt ions. In our most recent results, Fig. 4 shows that EO water can transform single crystal films of sodium cobalt oxide, which is made by lateral diffusion of sodium into $Co_3O_4$ (111) epitaxial film, to sodium cobalt oxyhydrates with bilayers of water molecules intercalated into the host lattice.

**Acknowledgements** This work is supported by the National Science Council of Taiwan, ROC, grant No. NSC 94-2112-M-018-001.

Figure 1 Powder x-ray diffraction patterns (Fe K$\alpha$ radiation) of γ-$Na_{0.7}CoO_2$ and EO water-treated $Na_x(H_2O)_yCoO_{2-\delta}$. All the reflection peaks for the latter are indexable based on the space group $P6_3/mmc$ and are the same as those of $Br_2$-treated superconducting $Na_{0.35}(H_2O)_{1.3}CoO_{2-\delta}$. The EO water-treated sodium cobalt oxyhydrate was obtained by immersing the γ-$Na_{0.7}CoO_2$ in EO water having ORP ~860 mV for 12 days. The EO water is directly flown into a beaker from a commercial EO water generator and used as an oxidizing and hydration agent.

Figure 2 Powder x-ray diffraction patterns (Fe K$\alpha$ radiation) of a) tap water-treated $Na_{0.29}(H_2O)_{0.41}CoO_{1.78}$ and (b) EO-treated sample being stored in the ambient environment for 21 days. A broad hump (*) centred at $2\theta \approx 23.6°$ is always found in the samples treated with tap water and EO water having insufficient ORP, which could be associated with the oxygen deficiency. A broader and asymmetric peak of (004) reflection indicates that the sample begins to show disordered sequence of water monolayers and bilayers between the $CoO_2$ layers. In our analyses of x-ray diffraction pattern for a mixture phase sample containing both the $c \approx 19.6$ Å and $c \approx 13.8$ Å phases, we found similar oscillation in the basal plane reflection linewidths (FWHM) with diffraction order as in the case of $Na_{1/3}(H_2O)_{1.5}TaS_2$.

Figure 3 Zero field cooled (open circles) and field cooled (filled circles) magnetization of superconducting $Na_x(H_2O)_yCoO_{2-\delta}$ prepared using electrolyzed/oxidized water. The magnetization data were measured in an applied field of 10 Oe using a superconducting quantum interference device (SQUID) magnetometer (Quantum Design).

Figure 4 X-ray diffraction patterns (Fe K$\alpha$ radiation) of the single crystal film of sodium cobalt oxide and the EO water-treated film of sodium cobalt oxyhydrate. The EO water-treated film was obtained by immersing it in the EO water for 8 days. The (002) reflection at $2\theta \approx 10.5°$ indicates the *c*-axis expansion due to the bilayers of water molecules intercalated between the $CoO_2$ layers.

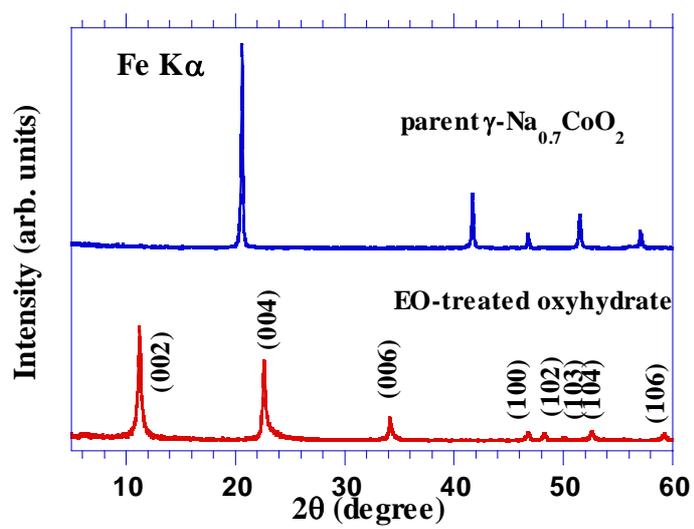

Fig. 1

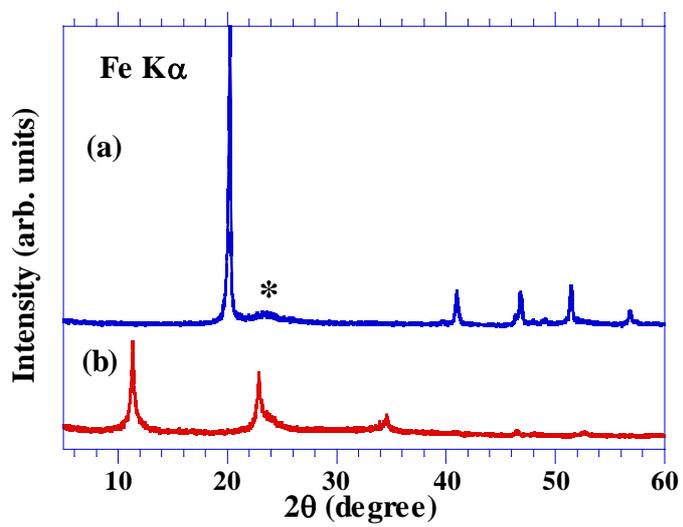

Fig. 2

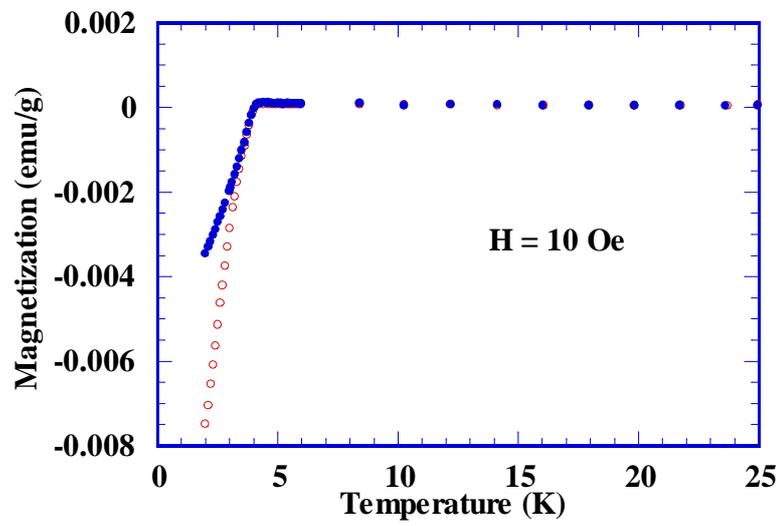

Fig. 3

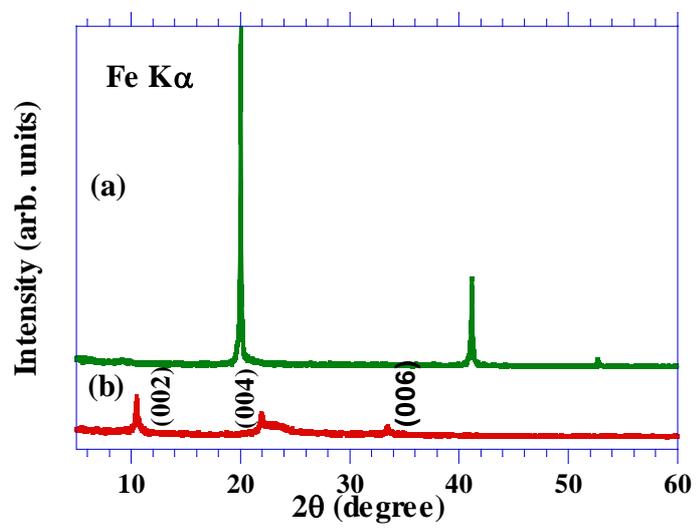

Fig. 4